\documentclass[usegraphicx,useAMS,usenatbib]{mn2e}

\usepackage{amssymb}
\usepackage{aas_macros} %needed to define journal macros generated by
\title[Multi-site campaign on M67. I.]{Multi-site campaign on the open cluster M67. I. Observations and photometric reductions} 
\author[D. Stello et al.]
{D. Stello$^{1,2,5}$,
T. Arentoft$^{1,2,8}$,
T. R. Bedding$^{2}$,
M. Y. Bouzid$^{4}$,
H. Bruntt$^{1,2,5}$,\newauthor
Z. Csubry$^{6}$,
Z.~E. Dind$^{2}$,
S.~Frandsen$^{1,8}$,
R. L. Gilliland$^{3}$,
A. P. Jacob$^{2}$,\newauthor
H. R. Jensen$^{1}$,
Y. B. Kang$^{7}$,
S.-L. Kim$^{7}$,
L.~L.~Kiss$^{2}$,
H. Kjeldsen$^{1,8}$,
J.-R.~Koo$^{7}$,\newauthor
J.-A.~Lee$^{7}$,
C.-U. Lee$^{7}$,
J. Nuspl$^{6}$,
C. Sterken$^{4}$ and
R.~Szab\'o$^{6}$\\
$^{1}$Institut for Fysik og Astronomi (IFA), Aarhus University, 8000 Aarhus,
Denmark\\
$^{2}$School of Physics, University of Sydney, NSW 2006, Australia\\
$^{3}$Space Telescope Sceince Institute, 3700 San Martin Dr., Baltimore, USA\\
$^{4}$Vrije Universiteit Brussel, Pleinlaan 2, B-1050 Brussels, Belgium\\
$^{5}$Department of Physics, US Air Force Academy, Colorado Springs, CO
80840, USA\\
$^{6}$Konkoly Observatory of the Hungarian Academy of Sciences, 1525 Budapest, PO Box 67, Hungary\\
$^{7}$Korea Astronomy and Space Science Institute, Daejeon 305-348, Korea\\
$^{8}$Danish AsteroSeismology Centre, Aarhus Universitet, DK-8000 Aarhus,
Danmark}

\begin{document}

\date{Accepted 2006 September 12. Received 2006 September 12; in original form
2006 June 06} 

\pagerange{\pageref{firstpage}--\pageref{lastpage}} \pubyear{2006}

\maketitle

\label{firstpage}

\begin{abstract}
%{\bf 
We report on an ambitious multi-site campaign aimed at detecting stellar
variability, particularly solar-like oscillations, in the red
giant stars in the open cluster M67 (NGC 2682). 
During the six-week observing run, which comprised 164 telescope nights,
we used nine 0.6-m to 2.1-m class telescopes located around the world 
to obtain uninterrupted time-series photometry.
We outline here the data acquisition and reduction, with emphasis on the
optimisation of the signal-to-noise of the low amplitude
(50--500$\,\mu$mag) solar-like oscillations.  
This includes a new and efficient method for obtaining the linearity
profile of the CCD response at ultra high precision ($\sim10\,$parts per
million). The noise in the final time series is $0.50\,\rmn{mmag}$ per
minute integration for the best site, while the noise in the Fourier
spectrum of all sites combined is 20$\,\mu$mag.  
In addition to the red giant stars, this data set proves to be very valuable
for studying high-amplitude variable stars such as eclipsing binaries,
W UMa systems and $\delta\,$Scuti stars. 
%} 
%evt lidt mere resultat
\end{abstract}


% {\it IRAS\/} <= LOOK AT THAT for italic

%use \hbox{60\,$\umu$m} <= LOOK AT THAT \hbox for no-brake and \umu for
%non-italic greek 

%\section[]{Description of the Envelope\\* Model}(Envelope\\* Model) <=
%LOOK AT THAT 

%\begin{equation}
%   L(\nu)=\mskip-12mu\int\limits_{\rmn{envelope}}
%   \mskip-12mu (\mskip-12mu) <= LOOK AT THAT
%   \rho(r)Q_{\rmn{abs}}(\nu)B[\nu,T_{\rmn{g}}(r)]
%   \exp [-\tau(\nu,r)]\>  (\>) <= LOOK AT THAT I think its a space marker
%\end{equation}

%${\balpha}$ for boldface greek

\section{Introduction}
%{\bf 
Asteroseismology of stellar clusters is potentially a powerful
tool. The assumption of a common age, distance, and chemical
composition provides stringent constraints on each cluster member, which
significantly improves the asteroseismic output
\citep{GoughNovotny93}. Hence, detecting  oscillations in 
cluster stars in a range of evolutionary states holds
promise of providing new tests of stellar evolution theory. 
Driven by this great potential, several studies have been aimed at 
detecting solar-like oscillations in the open cluster M67
\citep{GillilandBrown88,Gilliland91,GillilandBrown92b,Gilliland93} and in
the globular cluster M4 (Frandsen et al. in prep.). The most ambitious
campaign was reported by \citet{Gilliland93}, who used seven 2.5-m to 5-m
class telescopes during one week in a global photometric network to target 
11 turn-off stars in M67. Despite these efforts, they did not
claim unambiguous detection of oscillations. However, one of their
conclusions was that oscillations should be detectable in the more evolved
red giant stars due to higher expected oscillation amplitudes.
However, the oscillation periods of up to several hours and expected
frequency separations of a few $10^{-6}\,$Hertz would require a time base
of roughly one month on these stars. 
Recent month-long studies using single- or dual-site high-precision radial  
velocity measurements ($\sigma\sim2\,$m/s) on bright field stars have
clearly demonstrated that solar-like oscillations are present in red giant
stars \citep{Frandsen02,Barban04,Ridder06}.
However, due to non-continuous coverage these data suffered badly from
aliasing in the Fourier spectrum, which complicated the
detailed frequency analysis. Multi-site or space observations are
therefore required \citep{Stello06}. Such observations will hopefully soon
become available for red giant stars in the field from the MOST, COROT and
Kepler missions. 
However, after the cancellation of the ESA Eddington mission, no current or
planned space project will measure stellar oscillations in cluster
stars. Using high-precision spectrographs from ground to measure radial
velocities in red giants is not possible due to the lack of a global
network that can achieve high-precision velocities on an ensemble of
relatively faint cluster stars. Hence, the only feasible approach is
ground-based photometry.

\begin{figure}
 \includegraphics[width=84mm]{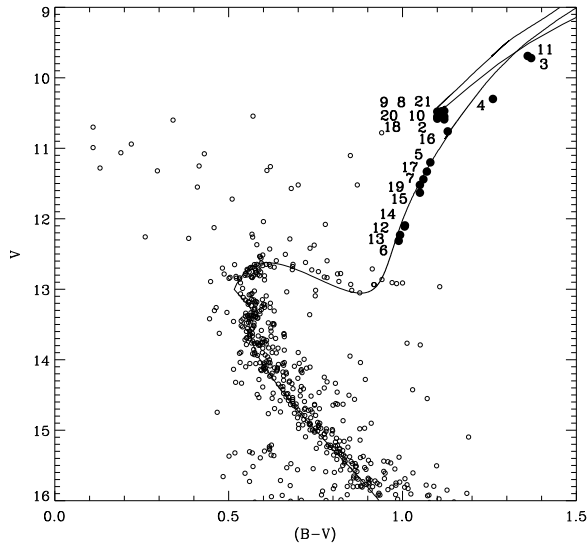}
 \caption{Colour-Magnitude diagram of the open cluster M67 (photometry by
 \citet{Montgomery93}). The red giant 
 target stars are indicated with filled symbols. The numbers  correspond to
 those indicated in  Fig.~\ref{fig2}. The solid line is an isochrone
 ($(m-M)=9.7\,\rmn{mag}$, $\rmn{Age}=4.0\,\rmn{Gyr}$,
 $Z=0.0198$ and $Y=0.2734$) from the BaSTI database 
 \citep{Pietrinferni04}.} 
\label{fig0}
\end{figure}

In this investigation we again target M67 using multi-site photometric
observations. Unlike the previous studies on this cluster, our primary
targets are the red giant stars (see Fig.~\ref{fig0}). 
%We have one target star in common with \citet{Gilliland93}
%one in common with \citet{GillilandBrown92} and seven targets are also in
%\citet{Gilliland91}. 
%Recent velocity measurements have established that red giant field stars do
%show solar-like oscillations, and 
Extrapolating the $L/M$-scaling relation \citep{KjeldsenBedding95} predicts
the amplitude of these stars to be in the range 50--500$\mu$mag. Although
very low, these amplitudes are significantly higher than for the turn-off
stars targeted by, e.g., \citet{Gilliland93}. 
%We note that recent 
%studies indicate that the $L/M$ scaling may overestimate the amplitude
%of dwarf and sub-giant stars, and that $(L/M)^{0.7}$ might provide a more
%realistic estimate \citep{Samadi05}. 
%Secondary targets are the stars that show variability with much higher
%amplitudes (3--400mmag). These include $\delta\,$Scuti stars and eclipsing
%binaries (see Fig.~\ref{fig0}). 
In addition to the red giants, more than 300 cluster stars were observed
during 
the campaign. Many are high-amplitude variables such as W UMa systems,
$\delta\,$Scuti stars and eclipsing binaries, some of which are already
known. This campaign provides a unique data set to investigate these
stars as well (Bruntt et al., in prep.). 

With emphasis on the low-amplitude red giant stars, the main purpose of
this paper is to describe the optimisation of time-series data towards
achieving the highest possible signal-to-noise in the Fourier spectrum (in
amplitude). Further discussion on the extraction of p-modes from the
Fourier spectra of these stars will be presented by Stello et al. (in
prep.).   
%Finally, using the known $\delta\,$Scuti star EW Cnc as an example, we want
%to illustrate what can be done with these data in terms of simple frequency
%analysis of high-amplitude variables.
%}

%We first report on the data acquisition and reduction, including the
%description of a new method for performing an efficient and
%ultra-high-precision linearity test of CCDs. Then we describe the weighting
%scheme optimised for the red giant stars, and evaluate our obtained noise
%with the estimated irreducible scatter.
%Finally, as an example of one of the high-amplitude variable stars we show
%the time series analysis of one of the $\delta\,$Scuti stars in the cluster.
%Results for the red giants will be reported in a second paper (Stello et
%al., in prep).

%In the following sections we describe the observations and data analysis
%optimised for the time series analysis of the red giant stars. As an exam
%of the data we show a few results of a delta sct ...

\section[]{Observations}
\begin{figure}
 \includegraphics[width=84mm]{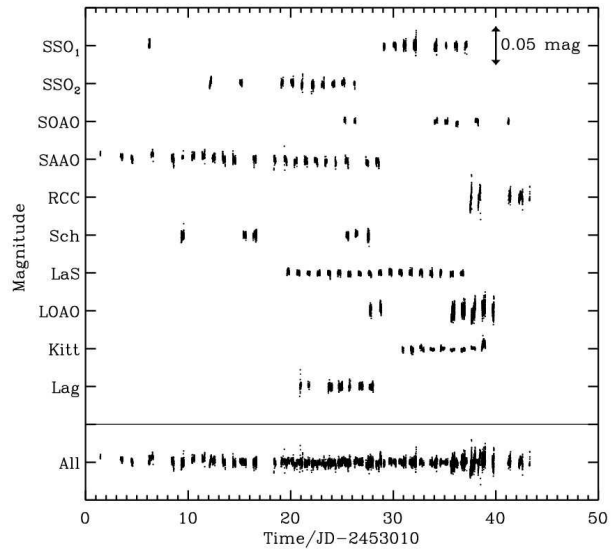}
 \caption{Time series of star No. 10 for all sites (after removing
          outliers and correcting for colour-dependent extinction; see
          Sect.~\ref{colour}). The site abbreviations are explained in
          Table~\ref{tab1}.}
\label{fig1}
\end{figure}

We observed the open cluster M67 from 6 January to 17 February 2004 using
nine telescopes (0.6-m to 2.1-m class) in a global multi-site network. 
The sites were distributed in longitude to allow continuous 
time series photometry during the six-week observing program. 
We were allocated 164 nights of telescope time which, due to bad weather,  
yielded about 100 clear nights (see Fig.~\ref{fig1} and
Table~\ref{tab1}). 
%{\bf 
In the first 18 days we observed 34\% of the time
and in the following three weeks the coverage was 80\%. For the
entire campaign (43 days) the coverage was 56\%.
%}

\begin{figure*}
 \includegraphics[width=120mm]{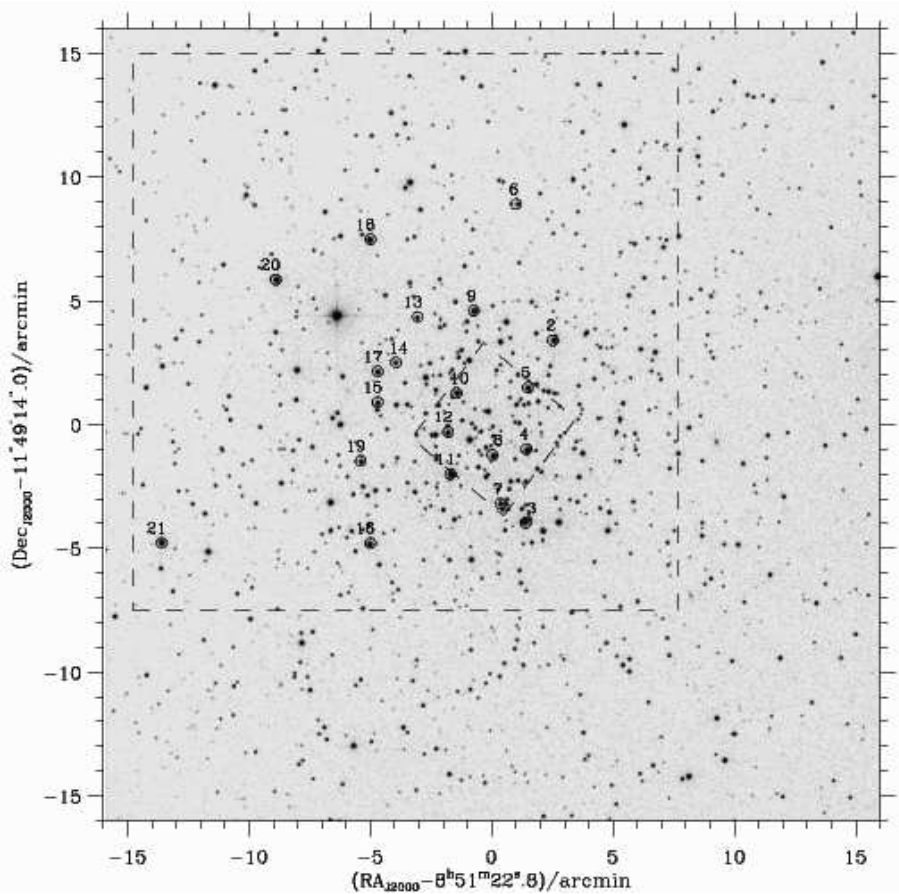}
 \caption{M67 field-of-view (FOV). SAAO with the smallest FOV and LOAO
          with the largest are indicated with dashed squares. The red
          giant stars are marked with circles. Source: STScI Digitized Sky
          Survey.} 
\label{fig2}
\end{figure*}

The participating telescopes and detectors had different properties and the
data sets are therefore rather diverse in terms of field-of-view (FOV), 
cadence and noise properties. A summary of the observations and the
instrument characteristics for each site is given in Table~\ref{tab1}.  
We indicate the smallest and largest FOV in Fig.~\ref{fig2}. The red giant
stars are indicated as well. %HR diagram for article with the science

%{\bf 
Observations at each telescope were planned to optimise the
signal-to-noise for solar-like oscillations in the red giant stars. We
did that by calculating both the noise and the expected
oscillation amplitudes (the signal) in different photometric filters. 
The amplitudes were estimated from \citet[~Eq.~5]{KjeldsenBedding95} and
the noise was estimated by photon counting statistics. These calculations
showed that the Johnson B and V filters were favourable, but an on-site
test was required to establish which of these was superior at each
telescope. The observers therefore chose filters based on
an initial test at the beginning of the first night. All sites except Kitt
Peak chose the V filter.
No phase change is seen in the solar oscillations between observations
obtained in different filters over the range 400--700 nm
\citep{Jimenez99}. We therefore expect the same adiabatic behaviour for
high-order solar-like oscillations in other stars as well. Hence, combining
data based 
on different filters can be done after a simple rescaling of the amplitude,
and corresponding adjustment of the weights to preserve the
signal-to-noise.
%}

\begin{figure}
 \includegraphics[width=84mm]{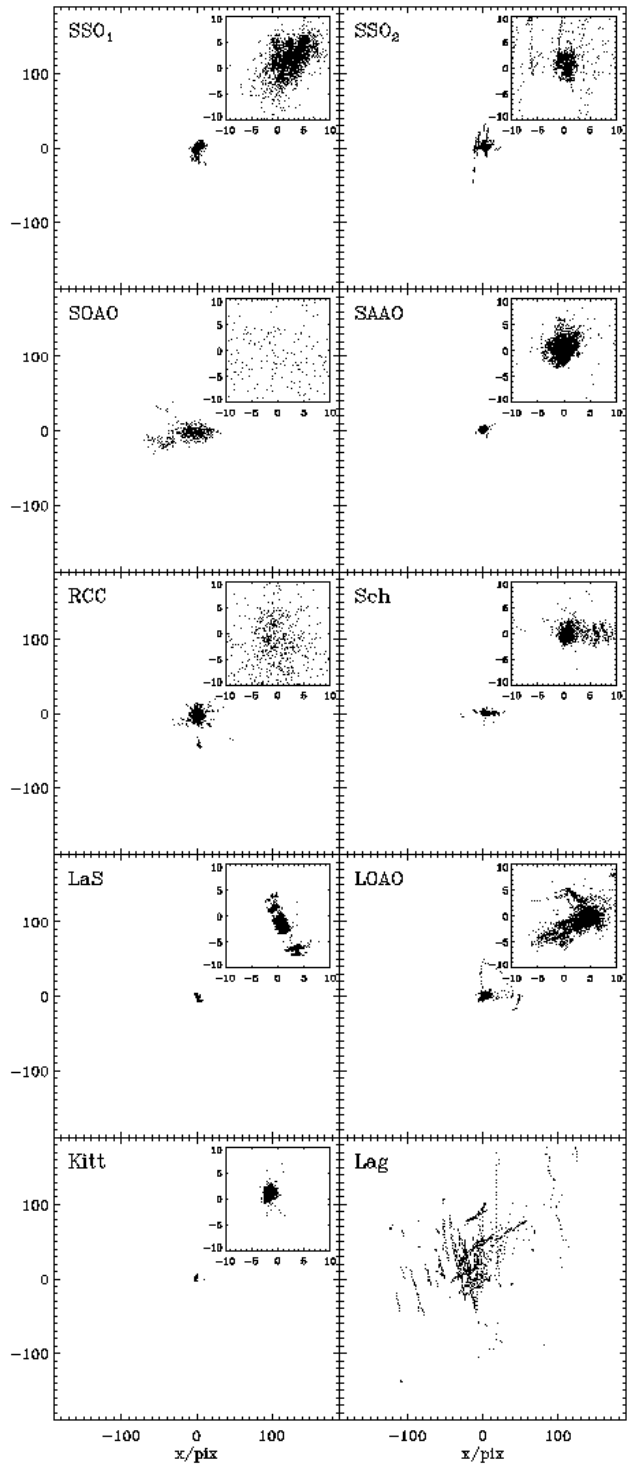}
\vspace{-1.5cm}
 \caption{The position on the CCD of star No. 10 relative to a reference
 frame. The insets show the inner 20 by 20 pixels. Although autoguiding is
 good at LaS, instrument rotation introduced 
 drift during the observing run for stars far from the rotation axis.} 
\label{fig1a}
\end{figure}

To obtain a noise level which was essentially limited by scintillation and
photon noise, it was important to avoid drift on the CCD of the stellar
field. The aim was to have each star confined within a few pixels. Not
all sites had autoguiding systems and as a result we found very 
different drift characteristics from site to site (see Fig.~\ref{fig1a}).
The images were defocused to obtain a higher duty cycle but we avoided
crowding.

\begin{table*}
 \centering
 \begin{minipage}{157mm}
  \caption{Summary of observations.}
  \begin{tabular}{@{}lrrrrrrrrrrr@{}}
  \hline
   Site\footnote{Site abbreviations and observers are: 
          SSO$_1$ (Wide Field Imager at Siding Spring Observatory,
          Australia, Z.E.D., D.S., A.P.J. and L.L.K.); 
          SSO$_2$ (Imager at Siding Spring, A.P.J., J.N. and D.S.); 
          SOAO (Sobaeksan Optical Astronomy Observatory, Korea, S.-L.K.,
          J.-A.L. and C.-U.L.);  
          SAAO (South Africa Astronomical Observatory, T.A. and H.R.J.); 
          RCC (Ritchey-Chr\'etien-Coud\'e at Piszk\'estet\H{o}, Konkoly
          Observatory, Hungary, J.N.);  
          Sch (Schmidt at Piszk\'estet\H{o}, Z.C. and R.S.); 
          LaS (La Silla Observatory, Chile, H.B.); 
          LOAO (Mt. Lemmon Optical Astronomy Observatory, Arizona,
          Y.B.K. and J.-R.K.);  
          Kitt (Kitt Peak National Observatory, Arizona, R.L.G.); 
          Lag (Mt. Laguna Observatory, California, C.S. and M.Y.B.).} 
 & Telescope & Filter & FOV    &  Image     & Observing & N$_{\rmn{exp}}$ & 
   Median  & Exp. & Duty & \#nights & \#nights \\%& Duty \\ 
 & aperture  &        &        &  scale     &  time     &                 &
   cadence & time & cycle&  alloc.  & good     \\%& cycle\\ 
 \hline
 SSO$_1$& 1.0m & V & 14\farcm0 & 0\farcs38/pix &  43.0h & 2205 &  62s &  30s & 48\%& 16 &  9 \\%& 56\%\\
 SSO$_2$& 1.0m & V & 14\farcm0 & 0\farcs60/pix &  51.0h & 1157 & 144s &  30s & 21\%& 17 & 10 \\%& 59\%\\
 SOAO   & 0.6m & V & 20\farcm5 & 0\farcs60/pix &  33.6h &  467 & 240s & 120s & 50\%& 17 &  7 \\%& 41\%\\
 SAAO   & 1.0m & V &  6\farcm0 & 0\farcs31/pix & 112.9h & 2595 & 149s &  80s & 54\%& 28 & 22 \\%& 79\%\\
 RCC    & 1.0m & V &  7\farcm0 & 0\farcs29/pix &  23.6h &  722 &  89s &  70s & 79\%& 14 &  5 \\%& 36\%\\
 Sch    & 0.6m & V & 17\farcm0 & 1\farcs10/pix &  31.3h & 1584 &  53s &  35s & 66\%& 16 &  6 \\%& 38\%\\
 LaS    & 1.5m & V & 13\farcm5 & 0\farcs39/pix & 109.0h & 3945 &  90s &  24s & 27\%& 22 & 18 \\%& 82\%\\
 LOAO   & 1.0m & V & 22\farcm5 & 0\farcs66/pix &  41.8h & 2886 &  46s &  12s & 26\%& 14 &  7 \\%& 50\%\\
 Kitt   & 2.1m & B & 10\farcm0 & 0\farcs30/pix &  75.5h & 1563 & 172s &  52s & 30\%& 11 &  9 \\%& 82\%\\
 Lag    & 1.0m & V & 14\farcm0 & 0\farcs41/pix &  46.4h & 1320 & 114s &  20s & 18\%&  9 &  7 \\%& 78\%\\
\hline
 Total  &      &   &           &               & 561.1h &18444 &      &      &     &164 &100 \\%& 61\%\\
\hline
\vspace{-1cm}
\label{tab1}
\end{tabular}
\end{minipage}
\end{table*}
%pt-pt er beregnet i ~/Idl/M67/Log/Figures/logMNRASpaper.txt til sidst
%sigma_amp er beregnet i ~/Idl/M67/Log/Figures/logMNRASpaper.txt sammen med
%fig sigma vs time


%evt erstat nedenstaaende par med:
%We adjusted the exposure time to have star No. 4 just below the saturation
%limit which implies that the two brighter stars No. 3 and 11 were often
%saturated. 

The exposure time at each site was adjusted to have star No. 4
just below the saturation limit, which provided a safety margin for the
large group of clump stars that were $0.3\,$mag fainter (see
Fig.~\ref{fig0}). However, the two brighter stars (No. 3 and 11) were
therefore often saturated. Due to their expected long oscillation periods
(Stello et al., in prep.), on time scales similar to typical instrumental 
drift, the results on these stars were likely to provide only limited
scientific output. Keeping these two stars above the saturation limit  
resulted in lower noise for the stars at the
base of the red giant branch (stars 6, 13, 12, 14) which were more likely
to produce useful results. 

\section[]{Calibration}
We calibrated each CCD image using four steps: 
\begin{enumerate}
 \item overscan subtraction (not all CCDs had an overscan region), 
 \item subtraction of bias (the bias levels were stable enough to use a
 single master bias image for each CCD),  
 \item correction for non-linearity (see Sect.~\ref{linearity}),
 \item flat-fielding to correct for pixel-to-pixel variations in the quantum 
efficiency (we used one master flat field for each CCD; for Kitt Peak and 
the RCC this was based on dome flats, while we used sky flats for all 
other sites).
\end{enumerate}
We found that the dark current was negligible compared to the
read-out-noise for all sites and it was therefore ignored.
These four steps were standard except the non-linearity correction, which 
is described in more detail in the following section. 

\subsection{CCD linearity calibrations}\label{linearity}
%We conducted both linearity tests at low and high light levels.
%Non-linearity at low light levels (also known as
%deferred charge or charge skimming) has been an issue for older 
%generation CCDs. Most sites acquired special low level linearity 
%images (similar to those described by \citet{Gilliland93}) 
%but the effects of deferred charge was not detected for any site. 

In this project a few target stars were relatively close to the CCD
saturation limit, at flux levels for which the non-linear response of 
the CCD gain could be significant. 
%TB suggest delete this sentence
%Although the CCD gain often varies 
%smoothly as a function of the light level and hence only affecting our 
%differential photometry very little, 
It is important to correct for these gain variations to attain the high
photometric precision required by this project.

\begin{figure}
 \includegraphics[width=84mm]{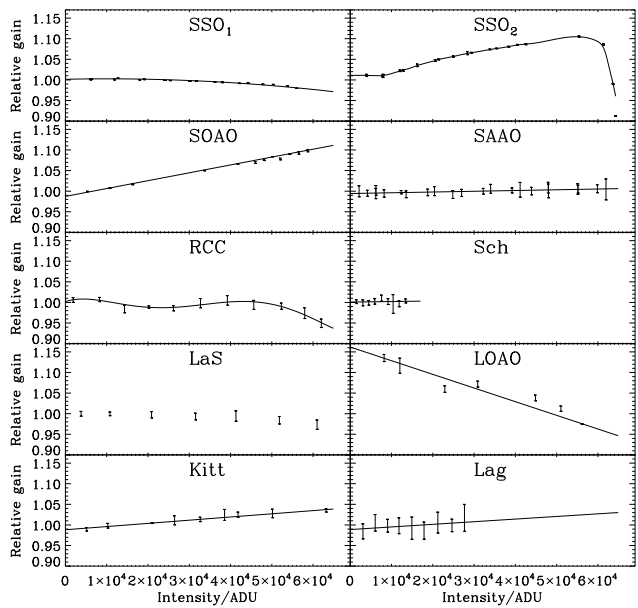}
 \caption{Classical linearity tests for all CCDs. For SSO$_1$, SSO$_2$ and
 SAAO two separate tests have been merged. Error bars are plotted as
 $3\times$rms to make them visible. 
 The solid lines are polynomial fits used to correct the data for
 non-linearity. 
} 
\label{fig2a}
\end{figure}

Linearity at high flux levels was investigated for all CCDs using a
``classical'' linearity test. We used an approach similar to that described
by \citet{Gilliland93}. 
The method measures relative variations in the CCD amplifier gain, rather
than absolute calibration in terms of $e^-$/ADU.
Flat-field images were taken sequentially with
increasing exposure time, interleaved with reference images; e.g. $3\,$s,
$10\,$s, $3\,$s, $10\,$s,$\cdots$, $3\,$s, then followed by $3\,$s,
$20\,$s, $3\,$s, $20\,$s,$\cdots$, $3\,$s,  
until the final series, in which the longer exposures were almost saturated.
The reference images allowed instabilities of the light source to be
measured and removed. In some cases, however, the flat-field lamp varied on
time scales too short to be sampled and a correction could therefore not be
made. The mean counts in the flat-field image, scaled according to the 
exposure time, were plotted versus the mean counts. Images with the same
exposure time were grouped to form a single point, with an uncertainty
equal to the group rms. In Fig.~\ref{fig2a} we show the results of the
linearity tests for all CCDs. Saturation occurred at $65536\,$ADU except for
the Schmidt, where it was at $16384\,$ADU. 
From Fig.~\ref{fig2a} (upper right panel) we see that non-linearity can
introduce variability in the stellar time series up to several percent in
non-photometric conditions (variable atmospheric transparency). 
Using ensemble photometry (Sect.~\ref{ensemble}) will, however, reduce this 
effect if the ensemble stars are roughly of equal colour and luminosity. 
We decided to correct for non-linearity for all sites except the Schmidt,
which did not show measurable non-linear effects, and Laguna, where we only 
had measurements of the 
gain variations up to approximately 30000 ADU, which was significantly lower
than the intensity levels of most target stars.
For LOAO, the data did not justify a description of the gain variation to be
higher than a first-order polynomial fit, although a few points could
indicate that higher-order features were present.
The linearity calibration of the data from La Silla was based on a 
method described in Sect.~\ref{newmethod}, hence no fit was applied to the
data shown in Fig.~\ref{fig2a}.

We obtained calibration images at La Silla for a new and more
elegant method for determining the linearity properties. This method
provides a much more precise determination of the CCD gain variations,
which we will compare with the classical method in the next section.

\subsubsection{Ultra-high-precision method}\label{newmethod}
\begin{figure}
 \includegraphics[width=84mm]{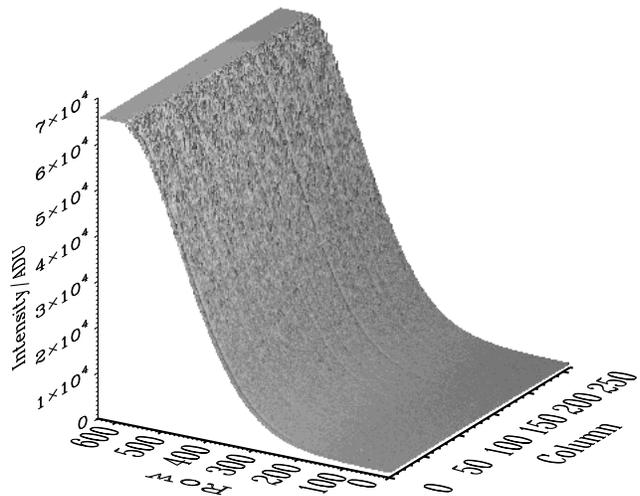}
 \caption{Linearity flat field of the longest exposure, $410\,$s (see
 text). The flat plateau at high row numbers is due to the digital
 saturation of the CCD.} 
\label{fig3}
\end{figure}

The basic concept of the method described in this section was first
outlined by \citet{Baldry99} and \citet{Knudsen00}, and was developed into
a fully applicable method by \citet{Stello02}.
Like the classical method, this method measures relative variations in
the CCD amplifier gain using  
flat-field images of different exposure times, but it differs by using 
flat fields that have a strong gradient e.g. by using a grism. 
Each flat field showed a large smooth variation in light level from
approximately the bias level to a significant fraction of the saturation
limit, with  the longest exposure reaching saturation (see
Fig.~\ref{fig3}).
The advantage of this method is that the effect from instabilities in the
light source used to obtain the flat fields is very small, because
we are sampling a large range (in the longest exposures 
the entire range) of the CCD gain response in a single exposure. The
resulting measurement  precision of the gain variations is several orders
of magnitude better than the classical method. 
Further, this method requires relatively few images to achieve high
precision, making it very efficient. 

\begin{figure}
 \includegraphics[width=84mm]{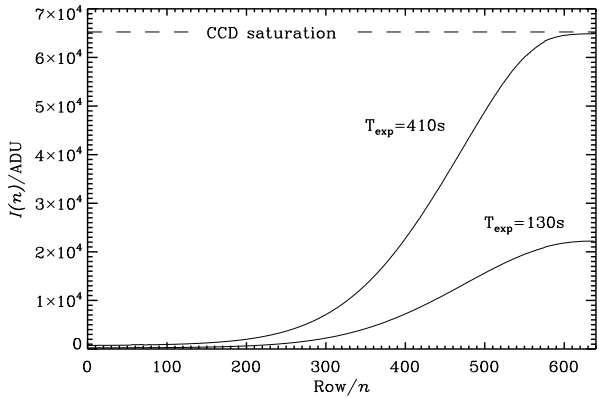}
 \caption{Two intensity curves of different exposure times, $410\,$s and
 $130\,$s (see text). The flat part of the $410\,$s exposure at high row
 number is due to saturation of the CCD.}
\label{fig4}
\end{figure}

We obtained spectral flats using the DFOSC spectrograph on the Danish 1.54-m 
telescope (La Silla). Light variation in one direction across the CCD was 
achieved using a grism to disperse the light from the slit illuminated by an 
internal telescope calibration lamp (Fig.~\ref{fig3}). 
A series of images were acquired in the following way: 
$3\times30\,$s, $3\times130\,$s,
$3\times30\,$s, $3\times250\,$s,
$3\times30\,$s, $3\times370\,$s,
$3\times30\,$s, $3\times390\,$s,
$3\times30\,$s, $3\times410\,$s, $3\times30\,$s. The control exposures of
$30\,$s enabled long-term drift in the flat field-lamp to be removed. 
Although this improves the precision, it is not critical.
After subtraction of overscan and bias, we corrected for the long-term
drift of the flat-field lamp and made an average (master) flat field for
each exposure time.
We then collapsed each master flat field by averaging in one
direction to obtain one dimensional \textit{intensity curves}, as shown in 
Fig.~\ref{fig4}.

Due to non-linear effects in the CCD, the intensity we measure in the 
$n$th row is:
\begin{eqnarray}
 I(n) &=& T_{\rmn{exp}} \cdot S(n) \cdot g(I(n)), 
\label{row_intens}
\end{eqnarray}
where $T_{\rmn{exp}}$ is the effective exposure time in seconds
(after correcting for dead-time of the shutter and short-term fluctuations 
in the light source), $g(I(n))$ is the CCD amplifier
gain as a function of the measured intensity
and $S$ defines the shape of the intensity curves, hence $S(n)$ expresses the
intensity in the $n$th row from a 1-s exposure.
Although the actual gain variations are a function of
the received flux, we assume the CCD amplifier has a well-defined
output signal for every input signal. 
%For simplicity we write $I(n)$ as $I$ in the follwing.

\begin{figure}
 \includegraphics[width=84mm]{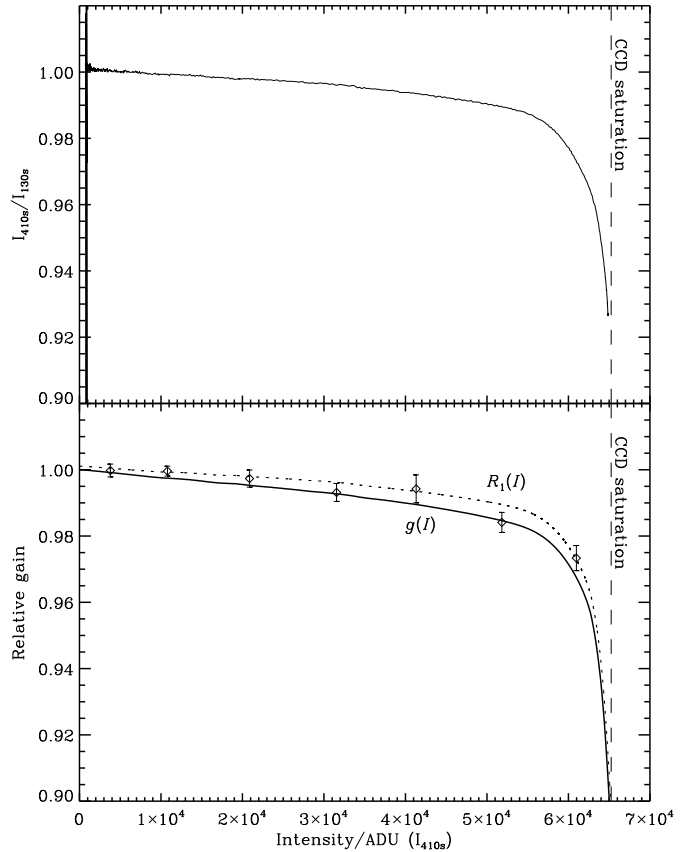}
 \caption{\textbf{Top panel:} Gain-ratio curve based on two intensity
 curves of different exposure time ($410\,$s and $130\,$s). \textbf{Bottom
 panel:} Inversion from smoothed gain-ratio curve $R_1(I)$ (dotted) to 
 final gain-curve $g(I)$ (solid). For comparison, the measurements from the
 classical method (Fig.~\ref{fig2a}, LaS) and their error bars ($1\times$
 rms) are indicated. The uncertainty on $g(I)$ is of the order $10\,$ppm. The
 smoothing of $R_1(I)$ introduces a systematic error in $g(I)$ of roughly
 1\% near the saturation limit because the curve is steep.}
\label{fig6}
\end{figure}

From two intensity curves with different exposure times, say $410\,$s 
and $130\,$s, we constructed the relative curve
\begin{equation}
 { I_{410\rmn{s}}(n) \over I_{130\rmn{s}}(n) } 
                         = { 410\,\rmn{s} \over 130\,\rmn{s} } \cdot 
                           { g(I_{410\rmn{s}}(n)) \over g(I_{130\rmn{s}}(n)) },
\label{rel_gain}
\end{equation}
\noindent
where we have corrected for shutter dead-time and verified that the shape 
$S$ was stable. 
%\begin{figure}
% \includegraphics[width=84mm]{fig5.eps}
% \caption{Gain-ratio curve based on two intensity curves of different
% exposure time ($410\,$s and $130\,$s).}
%\label{fig5}
%\end{figure}
In Fig.~\ref{fig6} (top panel) we show the curve
$I_{410\rmn{s}}(n)/I_{130\rmn{s}}(n)$ versus $I_{410\rmn{s}}(n)$, which we
call a \textit{gain-ratio curve}. At each intensity, $I_1$, this curve
shows the gain ratio, $g(I_1)/g(I_2)$, where
$I_2=I_1\cdot130\,\rmn{s}/410\,\rmn{s}$. Thus, finding $g(I)$ for any
intensity level requires inversion of the gain-ratio curve in an
iterative process.  

We started out using a smoothed version of the measured gain-ratio curve,
which we denoted $R_1(I)$, 
as a first estimate for the actual underlying gain curve $g(I)$ (see
Fig.~\ref{fig6}, bottom panel). Then, assuming $g_1(I)=R_1(I)$, we computed
a new 
gain-ratio curve, $R_2(I)=g_1(I)/g_1(I\cdot 130\,\rmn{s}/410\,\rmn{s})$ for
all $I$.  
The new estimate for the gain was then corrected by the
relative deviation between $R_{1}$ and $R_{2}$ according to
$g_2(I)={ R_{1} \over R_{2} }g_1(I)$, etc. This iterative process stopped
when $R_{i}$ matched the measured $R_{1}$ and the corresponding $g_i(I)$ was
the desired gain curve.
Before smoothing the measured gain-ratio curve (Fig.~\ref{fig6}, top panel)
we removed the noise at low 
intensities (0--1000 ADU) by replacing it with a linear fit to the data
points from 1000 to 10000 ADU.
%extrapolating down to 0 ADU from a straight line fit to the curve from 1000
%to 10000 ADU.
In the example shown, the gain-ratio curve is very similar to the final gain
curve. However, this is not a feature of the method, but is due to the 
gain characteristics of the particular CCD amplifier.

To ensure that all features in the gain were detected, we examined gain-ratio
curves based on different exposure-time ratios. If a feature, say a bump in
the CCD gain, is periodic for increasing intensity and hence repeated at
all pairs of intensities ($I_1,I_2$) related as
$I_2=I_1\cdot 410\,\rmn{s}/130\,\rmn{s}$, it will not 
show up in the gain-ratio curve based on $410\,$s and $130\,$s
exposures (or any combination with the same exposure-time ratio). Our gain
curves based on flat fields with different exposure-time ratios, all showed
an excellent match within the errors.  

In Fig.~\ref{fig6} (bottom panel) we compare our new method with results
from the classical method for La Silla (Fig.~\ref{fig2a}). The two methods
are in  agreement with each other, but the series of flat-field images for
our new method is significantly faster to 
obtain, provides the relative gain for all intensities and has a precision
more than 100 times better. However, it requires temporally
stable but spatially variable illumination of the CCD (e.g. spectral flat
fields) which is not possible at every telescope.

%We have tested the resulting gain curve (Fig.~\ref{fig6}, solid line) by
%correcting the $130\,\rmn{s}$ and $410\,\rmn{s}$ flat field images
%according to the gain curve. The gain curve resulting from these corrected
%flat field images is almost flat up to 62000 ADU
%

\section{Ensemble photometry}\label{ensemble}
The goal of this project is to measure relative light variations 
with very high precision. Hence, our approach is to obtain
differential photometry taking advantage of the ensemble of stars in the
FOV. Using an ensemble allows the effects from atmospheric variations,
common to all stars, to be removed from the time series.
The number of stars in the ensemble ranged from 116 (in the small FOV of
SAAO) to 358 (for LOAO).
We used the {\sevensize MOMF} photometry package \citep{KjeldsenFrandsen92}
to extract the photometry from the reduced images. It
calculates differential photometric time series by subtracting a
reference time series which is a weighted average based on all ensemble
stars. The weight given to each star is $1/\rmn{rms}^3$, which ensures that
stars with a high rms in their time series, such as faint stars and
high-amplitude variables, are strongly suppressed.  
{\sevensize MOMF} was developed especially to produce time-series photometry 
from large numbers of images (in particular defocused images) of
semi-crowded fields, similar to those obtained in this campaign. It
combines PSF and aperture photometry. 
We chose 10 stars, not necessarily the same for each site, to define the
shape of the point-spread-function (PSF). 
These were all non-crowded bright stars, i.e. mostly red giant stars
and a few bluer stars of the same luminosity (see Fig.~\ref{fig0}).
{\sevensize MOMF} allows multiple apertures and calculates the total rms,
$\sigma_{\rmn{total}}$, and the internal rms, $\sigma_{\rmn{internal}}$, of
the time series based on each aperture. The first is just the rms of the
time series while the latter is calculated as 
\begin{equation}
 \sigma_{\rmn{internal}}^2={1 \over 2(N-1)}\sum^{N-1}_{i=1}(m_i-m_{i+1})^2\,,
\label{int}
\end{equation}
\noindent
where $N$ is the number of points in the time series and $m_i$ is the
magnitude of the $i$th point in the series.
For each site and each star we chose the aperture with the lowest
$\sigma_{\rmn{total}}$ in the time series.

\section{Improving the photometry} \label{improve}
\subsection{Iterative sigma clipping}\label{sigmaclip}
\begin{figure}
 \includegraphics[width=84mm]{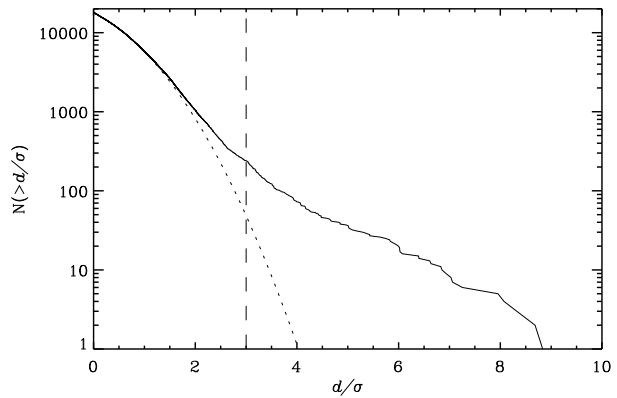}
 \caption{Cumulative distribution of $d/\sigma$ (solid line) for the entire
 data set of star No. 10. The dotted line is an analytical Gaussian
 distribution for comparison, and the dashed line indicates the threshold
 for our sigma clipping process.} 
\label{fig6a}
\end{figure}

\begin{figure}
 \includegraphics[width=84mm]{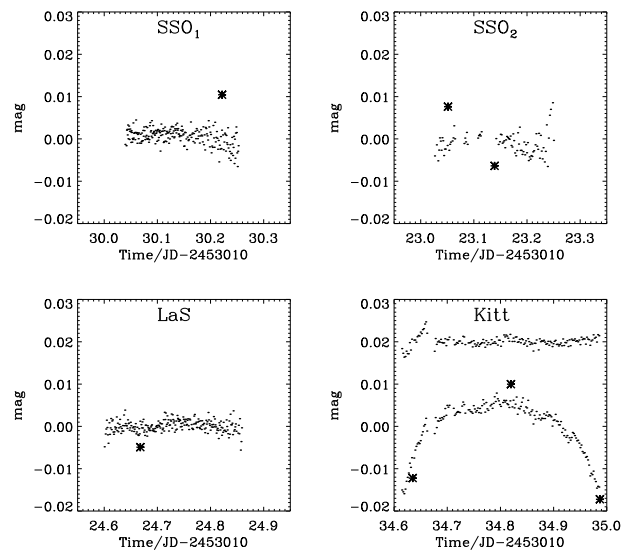}
 \caption{Photometric time series of star No. 10 for single nights from
 four sites. Outliers found by iterative sigma clipping are indicated with
 asterisks. The strong trend in the Kitt Peak data 
 %{\bf 
 (curved time series)
 %}
 is due to colour-dependent extinction not removed by the ensemble
 normalisation of {\sevensize MOMF}. The corrected time series, 
 %{\bf 
 has been
 %} 
 shifted upwards by 0.02 mag for clarity (see Sect.~\ref{colour}).}  
\label{fig7}
\end{figure}

To improve the overall quality of the data we first removed outliers. We
calculated the point-to-point deviation of each data point, $i$, relative to
its neighbouring points as 
\begin{equation}
 d_i  = m_i-0.5\cdot (m_{i+1}+m_{i-1})\,,
\label{d}
\end{equation}
\noindent
where $m$ is the magnitude. We then removed data points, $m_i$,
for which $d_i>3.0\sigma_i$, where $\sigma_i$ is the rms of $d$ 
within a 3-hour interval around data point $i$. This was done in an
iterative loop until no more points were removed, 
%{\bf 
which converged after a few iterations.
%} 
%{\bf 
We illustrate the underlying statistics of this process in
Fig.~\ref{fig6a}, which shows that the cumulative distribution of
$d/\sigma$ has a significant non-white tail of deviating data
points. Choosing the threshold is a trade-off between removing outliers and
keeping statistically valid points.  
The threshold was chosen so that roughly 85\% of the removed points were
real outliers and 15\% were valid points.
%}  
We tested our sigma clipping method on generated random noise to verify 
that it was not too drastic in terms of removing extreme points from pure
Gaussian noise. It removed less than 0.2\%, corresponding to about 35
points of the time series from the entire campaign. This should be compared
to the approximately 250 data points removed in total from the real
%{\bf 
time series for each star.
%} 
Results 
%{\bf 
of the sigma clipping on the
real data were further verified by visual inspection of diagrams
similar to what
%} 
is shown in Fig.~\ref{fig7}.   
%%%n addition, assuming our data is dominated by gaussian noise we compared
%%%he distribution of $(m_i-\langle m \rangle)/\sigma_i$ for each star with
%%%hat taken from computer generated gaussian noise, where $m$ is the
%%%hotometric time series. The data does not show any lack of extreme points
%%%elative to the gaussian distribution
%%%evt figur

\subsection{Colour extinction}\label{colour}
The data from Kitt Peak, which were the only ones obtained in the Johnson B
filter, showed clear residual trends. This arises from an uncorrected
wavelength dependence in the extinction (see Fig.~\ref{fig7}). 
These trends were visible in the red giant stars because the reference time
series calculated by {\sevensize MOMF} was dominated by bluer stars. 
Decorrelating the differential photometry against airmass still left a lot
of variation in the time series. Including more parameters, such as
sky background, in the decorrelation process would affect the expected
stellar signals significantly, which we verified with simulations.  
Subtracting a smoothed version of
the individual time series was also too harsh on the stellar signal.
Instead we tried using only the red giant stars themselves in the
ensemble. For some targets this provided good results but for most it did
not. 
%Simulations also showed that this approach can significantly
%affect the signal because the reference sample is relatively small.
Adding the much fainter main-sequence stars of similar colour to increase
the sample size did not improve the results. We finally chose to correct
the colour term from the {\sevensize MOMF} differential 
photometry in a similar way to \citet{GillilandBrown88}. For each image we
fitted a linear relation to the target stars:
\begin{equation}
 m_k - \langle m_k \rangle = a_0 + a_1\cdot (B-V)_k,
\end{equation}
\noindent
where $m_k$ is the magnitude of star $k$, $(B-V)_k$ is its colour and
$\langle m_k \rangle$ is the average of the time series. To correct star
$j$, we subtracted a fit that did not include the star itself
\begin{equation}
m_{\rmn{corr},j} = m_j - \langle m_j \rangle - (a_0 + a_1\cdot (B-V)_j).
\end{equation}
\noindent
This was to prevent stellar signal being removed by the process, which we have
confirmed with simulations. There were 12 red giant
target stars with low noise levels, hence 11 stars were used in each fit,
with a typical range of 1.00--1.25 mag in $B-V$ colour. 
In Fig.~\ref{fig7} (bottom right) we show the time series of one night of
star No. 10 before and after correction of the colour
term. The other sites showed weak effects from extinction, but these
trends were not consistent from star to star or night by night, and the
noise levels in the Fourier spectra did not improve if we performed the same 
correction as in the case of Kitt Peak. 
We therefore decided not to correct for residual extinction
in the differential photometry at any other site. 
%Instead, our stategy will be to remove peaks in the power spectra that
%occour at an integer 1--4 cycles per day as a result of these residual
%variations in the data (Stello et al. in prep.). 

\subsection{Weight calculation}\label{weight}
%The data is noise dominated but Fourier analysis will suppress the
%incoherent noise to a level that will allow us to detect coherent the
%stellar oscillations.
To be able to detect solar-like oscillations in the red giant stars, it is
crucial that we 
obtain noise levels as low as possible in the frequency range where the 
oscillations are expected to appear in the Fourier spectra of
the time series. 
%{\bf 
It is known that weighting time series of inhomogeneous data can
significantly improve the final signal-to-noise level
\citep{Handler03}. The important thing is that the
final weights represent the true variance of the noise on time
scales similar to the stellar signal one wants to detect. 
We will use a weighting scheme similar to that used by
\citet{Butler04} and \citet{Kjeldsen05} to minimise the noise in amplitude,
which includes the following two steps: 
\begin{enumerate}
\item Calculate weights from the point-to-point variance
  ($w_i=1/\sigma_i^2$).
\item Adjust the weights to obtain agreement between the noise at the
  relevant frequencies in the Fourier spectrum and the weights as being 
  represented by $w_i=1/\sigma_i^2$.
\end{enumerate}

(i) The point-to-point variance was not supplied by the photometric reduction
package and these values had to be estimated from the local variance of the
time series.
%} 
We estimated the local scatter $\sigma_i$ (=$\sqrt{\mathrm{variance}}$) for 
each data point $i$ as the rms of the $d$ array (Eq.~\ref{d}) 
%{\bf 
using a moving boxcar. The width of the boxcar (5 hours) was chosen to
minimise the noise (in amplitude) in the weighted Fourier
spectrum. The spectrum was calculated as a weighted discrete Fourier
Transform following the description of \citet{Frandsen95}.
%}  
%The effective width of the boxcar decreases towards the ends of each night. 
%Using $d$ to estimate the local scatter in the time
%series ensures that we are not sensitive to drift, because $d$ only samples
%the high frequency variations. 
Having first removed outliers, we prevented good data from being
down-weighted by bad neighbouring points during this process. 

%{\bf 
(ii) We then adjusted the weights night by night to be
consistent with the noise level (in amplitude), $\sigma_{\rmn{amp}}$,
between 300--900$\,\mu$Hz in the Fourier
%} 
spectrum, requiring that
$\sigma^2_{\rmn{amp}}\sum^N_{i=1}\sigma^{-2}_i=\pi$ 
\citep[Eq.~3 in][]{Butler04}. 
%{\bf 
The idea is that noise in this frequency range would have components
that affect the noise at slightly lower frequency as well where we expect
the stellar signal to be for the red giant stars. Choosing a frequency
range within the expected range of the stellar signal would effectively
down-weight stellar signal, which is not desired.
%} 

\begin{figure}
 \includegraphics[width=84mm]{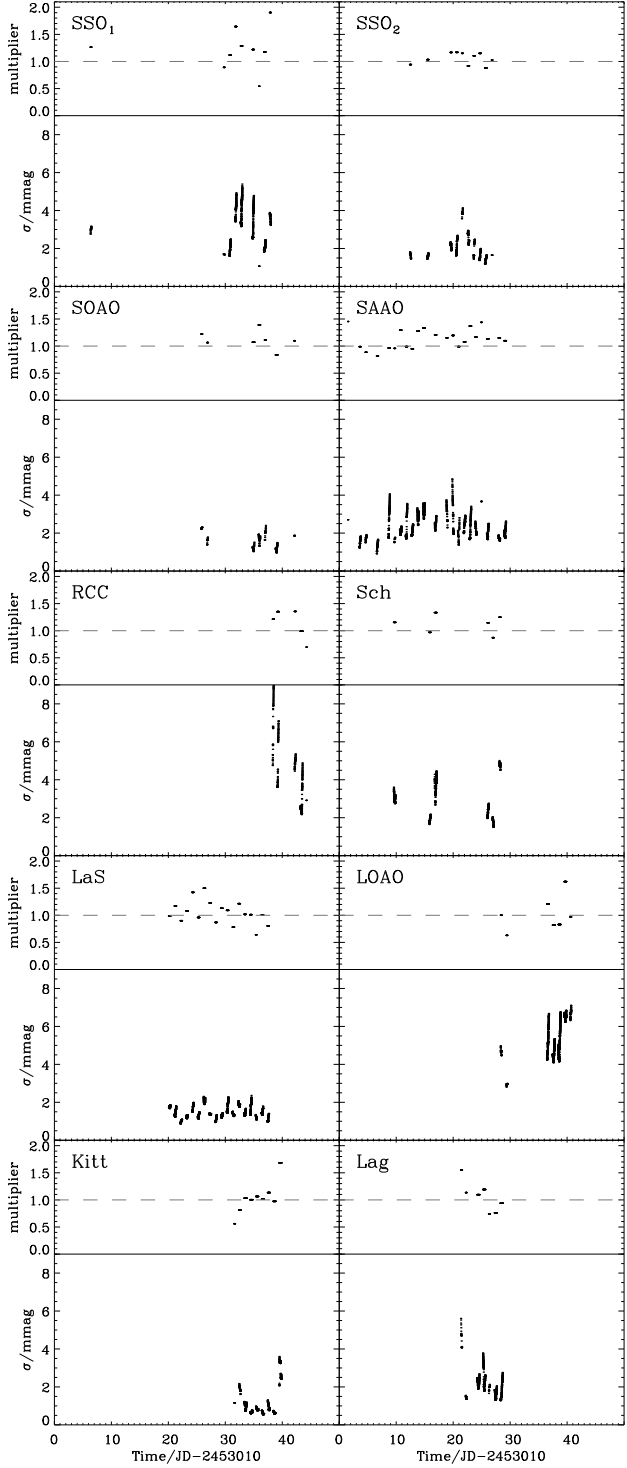}
 \caption{Scatter for each site for star No. 10. The
 insets show the multiplication factor used to adjust $\sigma_i$ for each
 night (see Sect.~\ref{weight}). The horizontal axes are the same as in the
 main panels.}   
\label{fig8}
\end{figure}

In Fig.~\ref{fig8} we plot our final estimates of $\sigma_i$, including the 
adjustment multipliers for each night shown in the insets. The maximum
adjustment was a factor of $\sim2$. For some sites, the noise 
in the final Fourier spectra in the range
300--900$\,\mu$Hz decreased by up to 20\% after adjusting weights on a
night-by-night basis, but in most cases it was a 5--10\% decrease. 
We see that $\sigma_i$ vary significantly during the observing run at
many sites. For example, the range at Kitt Peak is 0.54--$3.61\,$mmag (see
Fig.~\ref{fig8}). 
%{\bf 

\section{Error budget}
To establish whether the noise in the final time series was at
the irreducible lower limit dominated by photon noise and
atmospheric scintillation, we estimated each noise component and
compared with the measured noise in the time series 
%{\bf 
in a similar way
as in previous investigations by
\citet{GillilandBrown88,GillilandBrown92b} and \citet{Gilliland93}.
%}
%The telescopes and tht target stars in the \citet{Gilliland93} campaign were 
%different than in the present campaign and we therfore need to recalculate
%the error budget to be able to make a comparison between estimates and
%obtained results.
\begin{figure}
 \includegraphics[width=84mm]{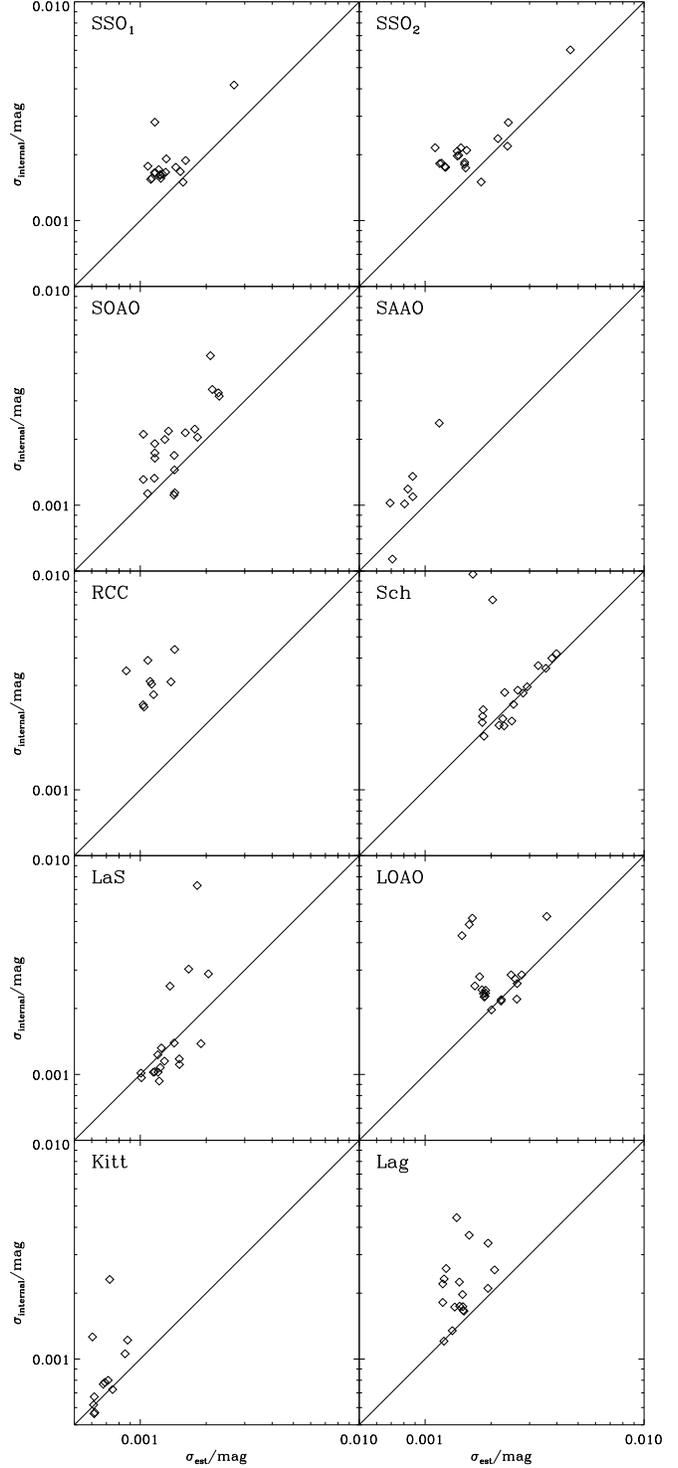}
\vspace{-1.5cm}
 \caption{Measured scatter versus estimated
   scatter for the red giant stars. The measured
   scatter is $\sigma_{\rmn{internal}}$ (Eq.~\ref{int}) of the relative
   photometry based on the best night for
   each star. The estimated  scatter is calculated from
   Eq.~\ref{estimate}, which gives relative errors, and is scaled by a
   factor of $1.086\,\rmn{ppm}/\mu\rmn{mag}$ to put it on the magnitude
   scale.}  
\label{fig9}
\end{figure}

Our total error budget comprised scintillation and counting statistics
within the aperture; the latter included stellar light, sky background,
level in the flat field, and CCD read-out-noise. 
The contribution from scintillation was estimated as
\begin{equation}
  \sigma^2_{\rmn{scint}} 
    = 0.09D^{-2/3}\chi^{3/2}\exp(-h/8000\,\rmn{m})T_{\rmn{exp}}^{-1/2}\,,
\end{equation}
\noindent
\citep[][~Eq.~3]{KjeldsenFrandsen92} using the factor of proportionality
from \citet{Young67}, $D$ the telescope diameter in centimetres,
$\chi$ the airmass, $h$ the elevation of the telescope in metres and
$T_{\rmn{exp}}$ the exposure time in seconds per image. 
For the counting statistics we used the expression from
\citet[][~Eq.~31]{KjeldsenFrandsen92}:
\begin{equation}
 \sigma^2_{\rmn{count}} 
    = {2\ln2 \over W^2\pi e_{\rmn{ff}}} + {1 \over e_{\rmn{star}}} 
     +\pi r^2_{\rmn{AP}} 
      {e_{\rmn{sky}}+\sigma^2_{\rmn{RON}} \over e^2_{\rmn{star}}}   \,,
\end{equation}
\noindent %evt ignore sig_ron
where $W$ is the full-width-at-half-maximum of the stellar
PSF in pixels, $e_{\rmn{ff}}$ is the number of electrons per
pixel in the flat field, $e_{\rmn{star}}$ is the number of electrons from
the star within the aperture, $r_{\rmn{AP}}$ is the radius in pixels of
the aperture, $e_{\rmn{sky}}$ is the number of electrons per pixel in 
the sky background and $\sigma_{\rmn{RON}}$ is the CCD read-out-noise (per 
pixel) in electrons. 
%{\bf 
Combining $\sigma_{\rmn{scint}}$ and $\sigma_{\rmn{count}}$ finally gives
the estimated scatter
%} 
\begin{equation}
 \sigma_{\rmn{est}} = (\sigma^2_{\rmn{scint}}+\sigma^2_{\rmn{count}})^{1/2}\,. 
\label{estimate}
\end{equation}
\noindent
%{\bf 
The estimated scatter was dominated by scintillation for the brighter
stars and by photon noise for the fainter stars. The 
magnitude at which the noise changed from being scintillation-dominated to
photon-noise-dominated was in the range $V=10.5$--$12.0\,$mag but different
from site to site. At a few sites the contribution to the 
counting statistics from the flat field was similar to the scintillation and
hence significant for the brighter stars. In general, the read-out-noise and
sky background could be neglected.
%}

%To obtain $\sigma_{\rmn{est}}$ we used the mean values of the actual
%parameters ($T_{\rmn{exp}}$,$W$,$e_{\rmn{star}}$ etc.) from each star during
%the chosen best nights. These values therefore vary from star to star. The
%jumps in the estimated scatter from star to star are mostly due to the
%different optimum aperture sizes in the photometry. The estimated scatter
%is a lower limit and any measured noise lower than that could be due to an
%overestimated CCD gain.

\begin{table*}
 \centering
 \begin{minipage}{140mm}
  \caption{Internal scatter $\sigma_{\rmn{internal}}$ in mmag of the red
    giant stars (sorted by their luminosity). The
    internal scatter is based on the entire time series using Eq.~\ref{int}
    after ensemble normalisation, sigma clipping and, for Kitt Peak,
    correction for colour-dependent extinction (see
    Sect.~\ref{improve}). Star No. 10 has the lowest scatter at all sites
    except SAAO and LOAO.}  
  \begin{tabular}{@{}rrrrrrrrrrrr@{}}
  \hline
No. & $m_V$ & SSO$_1$ & SSO$_2$ & SOAO & SAAO & RCC & Sch & LaS & LOAO & Kitt & Lag\\ 
 \hline
 3&  9.72 & 16.84& 23.69&  3.46&   --&   --& 13.11& 3.86&  8.18&   --&  7.75\\
11&  9.69 & 26.76& 10.28&  3.38& 4.11&   --& 16.81& 1.73& 10.10&   --&  2.88\\
 4& 10.30 &  2.37&  9.55&  1.87& 1.84& 5.16&  2.83& 1.39&  3.33& 2.77&  2.94\\
21& 10.47 &    --&    --& 20.63&   --&   --&    --&   --&  3.77&   --&    --\\
 8& 10.48 &  2.37&  8.13&  2.00& 1.70& 3.96&  2.58& 1.38&  2.95& 1.46&  2.51\\
 9& 10.48 &  2.73&  9.44&  1.81&   --&   --&  3.16& 1.42&  2.94& 1.22&  3.24\\
10& 10.55 &  2.35&  2.02&  1.44& 2.04& 3.87&  2.31& 1.36&  3.09& 1.11&  1.95\\
20& 10.55 &  2.86& 17.43&  2.85&   --&   --&  2.93& 1.58&  3.13&   --&  5.57\\
 2& 10.59 &  2.54&  4.83&  2.09&   --&   --&  2.91& 1.63&  2.92& 1.28&  4.09\\
18& 10.58 &  3.24& 16.04&  4.62&   --&   --&  3.12& 1.41&  3.07&   --&  7.64\\
16& 10.76 &  2.96&  9.69&  3.16&   --&   --&  3.28& 1.55&  7.93&   --&  5.15\\
 5& 11.20 &  2.44&  5.22&  3.04& 1.82& 4.68&  3.68& 1.96&  2.63& 1.62&  2.78\\
17& 11.33 &  2.72&  5.25&  2.47&   --& 5.66&  3.22& 3.50&  2.45& 1.91&  3.20\\
 7& 11.44 &  2.83& 13.65&  2.83& 2.00& 6.73&  3.31& 1.54&  2.57& 1.61& 17.38\\
19& 11.52 &  2.91&  4.86&  2.85&   --&   --&  3.20& 1.64&  2.95& 4.81&  3.60\\
15& 11.63 &  2.50&  4.54&  3.42&   --& 5.33&  3.54& 6.22&  2.66& 1.83&  2.88\\
14& 12.09 &  3.02&  5.40&  3.96&   --& 4.75&  4.22& 1.66&  3.23& 3.59&  2.90\\
12& 12.11 &  5.39&  8.88&  8.47& 3.98& 8.19&  4.30& 8.36&  6.04&   --&  9.02\\
13& 12.23 &  3.49&  8.74&  5.41&   --&   --&  4.75& 3.72&  3.32& 3.53&  3.26\\
 6& 12.31 &    --& 13.16&  5.02&   --&   --&  6.24&   --&  3.47&   --&    --\\
\hline
%\vspace{-1cm}
\label{tab3}
\end{tabular}
\end{minipage}
\end{table*}

In Table~\ref{tab3} we give the measured internal scatter (Eq.~\ref{int})
for each red giant star based on the full time series, which shows the
overall quality of the data from star to star and from site to site. In
general, star No. 10 had the lowest noise except for SAAO and LOAO.
%{\bf 
To compare with the estimated scatter (Eq.~\ref{estimate}) we
have, for each star and each site, measured the internal scatter
(Eq.~\ref{int}) for the best night, and the results are shown in
Fig.~\ref{fig9}.
There are other sources of noise not included in our error budget, which
can explain why some stars fall significantly above the line of
proportionality.
%}  
Saturation of the CCD will increase the noise significantly. For several
sites, stars No. 3, 4 and 11 were affected by saturation, which explains
their higher noise levels.
Close neighbouring stars can introduce higher noise in the photometry.
Star No. 12 has three close neighbouring stars and at most sites it suffers
from excess noise. If a star is located close to a bad column on the CCD
the noise will also increase, which we see in some cases.

In summary, for stars that are not affected by crowding, saturation or bad
columns, we generally see noise levels limited by photon and scintillation
noise on the best nights. At one site (RCC) the noise is larger than
estimated by a factor of two, which is unexplained.

\subsection{Noise comparison with previous campaigns}
The first campaign aimed at detecting solar-like oscillations in M67 
was carried out by \citet{GillilandBrown88}, who used a 0.9-m telescope
on two nights. Noise levels to $\sim1.5\,$mmag per minute integration were
attained for non-saturated stars ($m_V \ga 12\,$mag). 
%Paper II: but it did not provide any detection of oscillations.  
Later, \citet{Gilliland91} observed M67 for two weeks from five sites using
0.6-m to 1.1-m class telescopes. The lowest rms in the time series of the
non-saturated stars ($m_V \ga 12.5$) was $0.88\,\rmn{mmag}$ per minute
integration after high-pass filtering the data. 
%Paper II: but no unambiguous detections were claimed here either.
%Til paper II
%Upper limts of 100-150$\,\mu$mag for the amplitudes were given.
For comparison, our best 0.6-m site (SOAO) showed an internal scatter
(comparable to the rms of high-pass filtered data) down to
$\sim 1.6\,$mmag per minute integration, and
the best 1.0-m site (SAAO) showed $\sim 1.2\,$mmag per minute integration. 
A final, but direct, comparison can be made between our Kitt Peak
observations of the red giant No. 10 with those obtained by
\citet{GillilandBrown92b} using the same telescope in a similar campaign.
Star No. 10 \citep[No. 7 in][]{Gilliland91} was one of their targets with the
lowest noise, which was $0.43\,$mmag per minute integration after high-pass
filtering and decorrelating the data.
In comparison we obtained $0.50\,$mmag (based on the internal scatter, but
without any high-pass filtering or decorrelation).

To summarise, we find noise levels as good as in previous
studies based on similar size telescopes. However, this campaign has
provided significantly longer time series (six weeks compared to a maximum
of two weeks) with better coverage than earlier comparable campaigns,
%{\bf
which implies lower noise levels in the final Fourier spectra.
%}

%rms of $0.77\,$mmag after high-pass filtering and decorrelating the
%data. In comparison find an rms error of $0.75\,$mmag, which is without
%high-pass filtering or decorrelation 
%againts external parameters. Our internal error, however, is only
%$0.54\,$mmag for that star. If we compare the noise per minute integration
%the results are very similar though, with $0.43\,$mmag
%\citep{GillilandBrown92b} and  $0.50\,$mmag (this study), respectively.
%This comparison should be for the next paper (amp spec anal)
%\citet{Gilliland93} obtained a rms of
%$\sim0.25\,$mmag for star No. 7 (G12) with a $4m\,$ class telescope after
%high-pass filtering and decorrelating the time series. In comparison we
%obtain, for the same star, an internal error of $0.80\,$mmag with a $2.1m\,$
%telescope. 

%\section{Evt decorr of andet timeseies anal i denne paper}
%Although we ended not using the decorr for the data presented in this paper
%the method can be very powerful especially if automated. We therfore present
%it here:
%Because the stellar signal we want to measure varies on time scales
%comparable to the non-stellar drift in the data all steps
%enforced/performed on the data to reduce the noise have to be done with
%caution. In many cases it is better not to correct the data as the
%correction effects might lead to ambiguous results.

\subsection{Noise in the Fourier spectra}
The noise levels in the Fourier spectra obtained by our weighting
scheme were 20--30$\,\mu$mag for the red giant stars that were
not affected by excess noise from crowding, saturation or bad pixels.    
We therefore expect to be able to detect oscillations 
in the Fourier spectrum with $\rmn{S/N}\gid4$.  
However, the detection threshold depends very much on the mode lifetime,
which is unknown for these stars \citep{Stello06}, and will require
extensive simulations to quantify. This analysis will be published in a
subsequent paper. 

\begin{figure}
 \includegraphics[width=84mm]{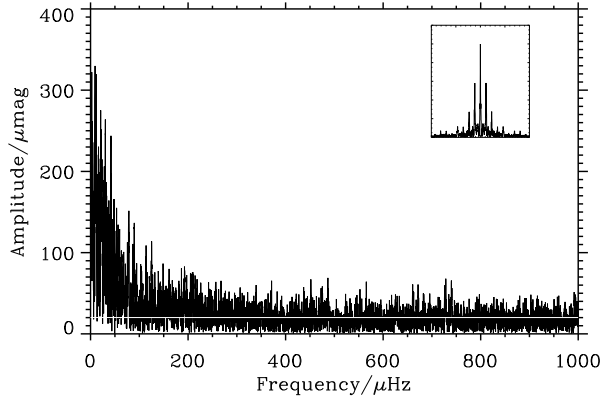}
 \caption{Fourier spectrum of star No. 10 (in amplitude). The white line
 indicates the noise level in the range 300-900$\,\mu$Hz. The inset shows
 the spectral window, which is on the same frequency scale as the main
 panel. Each data point has been weighted according to the 
 weighting scheme described in Sect.~\ref{weight}.}   
\label{fig11}
\end{figure}

In Fig.~\ref{fig11} we show the Fourier spectrum of star No. 10, which is
one of the best, to illustrate the noise level we have obtained. The noise
level, indicated with the white line, is the average in the range
300-900$\,\mu$Hz.
%, which 
%is just outside the frequency range were we expect the stellar oscillations. 
%Although, this example shows no clear evidence of oscillations, we do see
%excess power for some of the other red giants. 
The detailed pulsation analysis of all red giants will be presented by
Stello et al. (in prep.). 

\section{Conclusions}
%We have obtained point-to-point scatter in time series photometry
%throughout the entire observing run of
%1.1--2.5$\,$mmag as expected for all sites but one (0.6-$2.1\,$m class
%telescopes).   
%{\bf 
We have collected 100 telescope nights of photometric multi-site data of
the open stellar cluster M67 over a six-week period. 
The focus of this paper was the discussion of our approach towards
achieving the highest signal-to-noise ratio for the very low amplitude 
(50--500$\,\mu$mag) solar-like oscillations in the red giant stars. 
This included our careful reduction of the CCD images to
obtain the lowest possible noise in the time-series data, and our weighting 
scheme to reduce the noise level in the Fourier spectrum.

%Our main aim is
%to use these data to detect solar-like oscillations in the red giant stars
%in the cluster -- the results of which will be presented in a subsequent
%paper (Stello et al., in prep). The oscillation amplitudes are expected to
%be very small (50--500$\,\mu$mag) and will require a very high photometric
%precision to detect.

We have obtained a point-to-point scatter in the time-series photometry
down to about 1$\,$mmag for most sites, while 
the largest participating telescope reached 0.5$\,$mmag (Fig.~\ref{fig9}).
These values are similar to those from previous campaigns on M67 by
\citet{GillilandBrown88}, \citet{Gilliland91} and \citet{GillilandBrown92a}
which all used telescopes of similar size but for shorter time spans
(maximum two weeks).  
Comparison of our best nights with known noise terms 
demonstrates that the attained point-to-point
scatter is consistent with irreducible terms dominated by photon and
scintillation noise for all sites but one (RCC), which shows extra noise 
of unknown origin (Fig.~\ref{fig9}). 
With these scatter values, our weighting scheme provided a mean noise level 
in the Fourier spectra of approximately 20$\,\mu$mag (in amplitude), which
would allow us to detect solar-like oscillations in the red giant stars
with $\rmn{S/N}\gid4$ assuming $L/M$-scaling. 
%However, due to their low expected oscillation frequencies 
%(10--$300\,\mu$Hz) we will be very sensitive to drift in the data,
%hence the long-term stability and homogeneity of the data set will
%be the limiting factor.

%Finally, to illustrate the value of this data set towards studying
%classical variable stars, we have reported 33 oscillation frequencies
%(22 with $\rmn{S/N>4}$) in the blue straggler EW Cnc (S1280), which is a 
%$\delta\,$Scuti-type variable. This more than triples the number of known
%frequencies for this star.
%%The low sidelobes in the spectral
%%window make it unlikely that these frequencies are shifted by one daily
%%alias ($11.57\,\mu$Hz).
%We confirm 9 of the 10 frequencies previously detected
%\citep{GillilandBrown92b}, the tenth possibly alias shifted by
%$11.57\,\mu$Hz. This result promises great prospects for using this data
%set on other large-amplitude variables ($\delta\,$Scuti stars, eclipsing
%binaries and W UMa systems) in this cluster (work in progress).
%%} 

\section*{Acknowledgments}
This work was partly supported by the IAP P5/36 Interuniversity Attraction
Poles Programme of the Belgian Federal Office of Scientific, Technical and
Cultural Affairs.
This paper uses observations made from the South African Astronomical
Observatory (SAAO), Siding Spring Observatory (SSO) and the Danish 1.5m
telescope at ESO, La Silla, Chile. 
This research was supported by the Danish Natural Science Research Council
through its centre for Ground-Based Observational Astronomy, IJAF.

\bibliography{bib_complete}

\end{document}